\documentclass[epj]{svjour}
\usepackage{graphicx,color}

\begin{document}
\title{Light cluster production at NICA}
\author{N.-U. Bastian\inst{1}, 
P. Batyuk\inst{2}, 
D.~Blaschke\inst{1,2,3}, 
P.~Danielewicz\inst{4}, 
Yu.~B.~Ivanov\inst{3,5},
Iu.~Karpenko\inst{6,7}, 
G.~R\"opke\inst{3,8,}\thanks{\email roepke@uni-rostock.de},\\
O. Rogachevsky\inst{2},
H.~H.~Wolter\inst{9}
}                     
\titlerunning{Light cluster production at NICA}
\authorrunning{
N.-U. Bastian et al.}
%
%
\institute{
$^1$University of Wroclaw, 50-024 Wroclaw, Poland\\
$^2$Joint Institute for Nuclear Research, Dubna 141980, Russia\\
$^3$National Research Nuclear University (MEPhI), 115409 Moscow, Russia\\
$^4$Michigan State University, East Lansing, MI, USA\\
$^5$National Research Centre "Kurchatov Institute", 123182 Moscow, Russia\\
$^6$Bogolyubov Institute for Theoretical Physics, 03680 Kiev, Ukraine\\
$^7$INFN - Sezione di Firenze, I-50019 Sesto Fiorentino (Firenze), Italy\\
$^8$University of Rostock, D-18051 Rostock, Germany\\
$^9$Universit\"at M\"unchen, D-85748 Garching, Germany 
}
\date{Received: date / Revised version: date}
%
\abstract{
Light cluster production at the NICA accelerator complex offers unique possibilities to use these states as
"rare probes" of in-medium characteristics such as phase space occupation and early flow.
In order to explain this statement, in this contribution theoretical considerations from the nuclear statistical equilibrium model and from a quantum statistical model of cluster production are supplemented with a discussion of a transport model for light cluster formation and with results from hydrodynamic simulations combined with the coalescence model.  
} 
\maketitle
\section{Introduction}
To decide on characteristics of the matter produced in energetic nuclear collisions a variety of probes is needed that provide different perspectives.
One class of such probes are nuclear clusters.
Here we discuss the insights that the clusters can provide within the NICA program, both within fixed 
target and collider setups, at BM@N and MPD, respectively. 
 
\begin{figure}[!t]
\resizebox{0.5\textwidth}{!}{%
\includegraphics{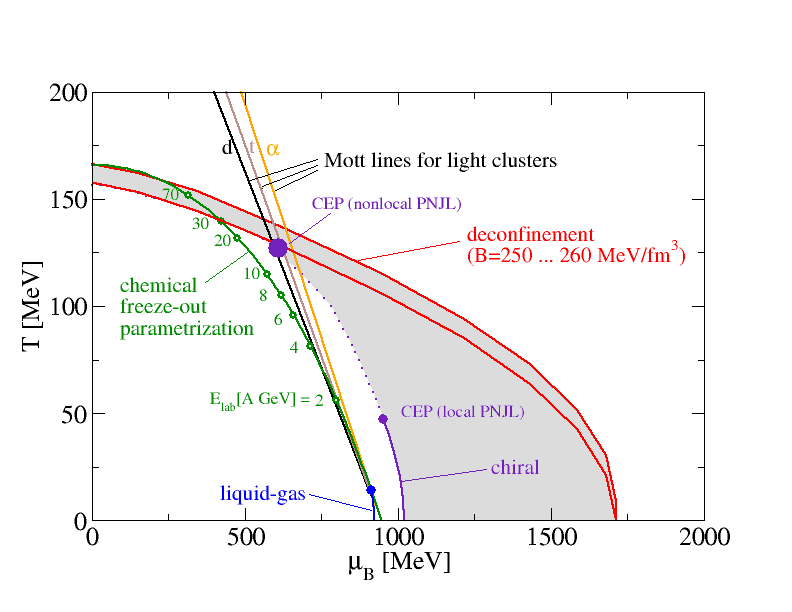}
}
\caption{Phase diagram of dense nuclear matter in the plane of temperature $T$ and baryochemical potential $\mu_B$.
The diagram includes Mott-lines for the dissociation of light nuclear clusters, extrapolated also to the deconfinement region. For details, see text.
\label{PD-Mott}      
}
\end{figure}

To provide a general background we show in Fig.~\ref{PD-Mott} a phase diagram of dense nuclear matter including lines for Mott dissociation of deuterons ($d$), tritons ($t$) and alpha particles 
($\alpha$), taken from \cite{Schuck:1999eg}, see also \cite{Roepke:1983},  together with the parametrization of the chemical freeze-out line \cite{Begun:2012rf} from statistical model fits of hadron production in heavy-ion collisions. 
Several laboratory energies from the energy range accessible at the 
NICA accelerator complex are shown as labeled dots on that line.

For the sake of generating the illustration, the hadronic phase is described by a DD2 equation of state 
\cite{Typel:2009sy}
with a liquid-gas phase transition (blue line with critical endpoint), extended by adding components of the hadron resonance gas, in particular pions and kaons.
The quark-gluon matter phase for the figure is described by a PNJL model exhibiting a first order/crossover chiral transition with a critical endpoint position that can be adjusted with details of the model (local or nonlocal, see \cite{Contrera:2012wj,Contrera:2016rqj}). 
The deconfinement lines are obtained requiring the baryon chemical potential and pressure in the DD2 hadronic and PNJL quark matter to be equal. 
The latter pressure is shifted by a (gluonic) bag constant of either 250 or 260 MeV/fm$^3$, respectively.

\section{Light clusters in nuclear matter}

\subsection{Law of mass action}

If the freeze-out model, describing hadron abundances in terms of specific temperature and chemical potential values, is extended to nuclear clusters, the law of mass action (LMA) \cite{DasGupta:1981xx} 
applies to cluster abundancies.
Under LHC conditions, the freeze-out model 
\cite{Andronic:2010qu,Cleymans:2011pe,Andronic:2011yq}
appears to describe the abundancies very accurately
\cite{Adam:2015vda}.
Very recently, the data for production of light nuclei in the NA49 experiment at SPS have been published which are interpreted using the statistical model too \cite{Anticic:2016ckv}.  
  
In terms of laboratory energy per nucleon $E_{\rm lab}$, for fixed target, the freeze-out curve can be approximated with \cite{Begun:2012rf}

\begin{eqnarray}
\label{freezeout}
\frac{T}{{\rm GeV}} &=& 0.166 - 0.139~\left(\frac{\mu_B}{{\rm GeV}}\right)^2 
                                         - 0.053~\left(\frac{\mu_B}{{\rm GeV}}\right)^4 ~,\nonumber\\
\frac{\mu_B}{{\rm GeV}} &=& \frac{1.308}{1+0.273\sqrt{s_{NN}}/{\rm GeV}}~,\\
\nonumber\\
\sqrt{s_{NN}}&=&\sqrt{2m_N E_{\rm lab} +2m_N^2}~;~m_N=0.939~{\rm GeV}~.\nonumber
\end{eqnarray}

Within the LMA description 
protons $p$, neutrons $n$ and their
clusters $d$, $t$, $^3$He ($h$) and $\alpha$ are considered as pointlike, noninteracting particles.
Density and pressure for each species $i$ of pointlike particles (superscript $"{\rm pt}"$)
are given by
\begin{eqnarray*}
	n^\mathrm{pt}_i (T, \mu_i) &=& \frac{g_i}{2\pi^2} \int_0^\infty \mathrm dk k^2 f(k) 
	\frac{1}{\mathrm e^{(E_i - \mu_i)/T} \pm 1}~,
\\	p^\mathrm{pt}_i (T, \mu_i) &=& \frac{g_i}{2\pi^2} \int_0^\infty \mathrm dk k^2 f(k) 
\frac{k^2}{3E_i}\frac{1}{\mathrm e^{(E_i - \mu_i)/T} \pm 1}~,
\end{eqnarray*}
respectively, where the $\pm$ signs refer to the appropriate statistics and $E_i=\sqrt{k^2+m_i^2}$.
The function $f(k) = 1 - \frac{3\pi}{4kR} + \frac{1}{(kR)^2}$ 
is a correction due to the finite size of a spherical freeze-out configuration with the radius $R$ 
\cite{BraunMunzinger:1995bp}.

In a next approximation step, one can account for effects of finite "hard core" radii $r_i=1.1~A_i^{1/3}$ fm 
of the clusters (with mass number $A_i=Z_i+N_i$, mass $m_i=Z_i m_\mathrm p + N_i m_\mathrm n + B_i$, chemical potential $\mu_i=A_i \mu_B-B_i$ and binding energy $B_i$) which result in an excluded volume 
$v^\mathrm{ex}_{ij} = \mathrm{2\pi} (r_i + r_j)^3/{3}$ .
Thermodynamic consistency is achieved in the standard way by shifting the chemical potential 
\begin{eqnarray*}
	\mu_i \rightarrow \tilde\mu_i &=& \mu_i - \sum_j p_j v^\mathrm{ex}_{ij}~,
\end{eqnarray*}
so that pressure and particle densities are to be determined selfconsistently from
\begin{eqnarray*}
 p (T, \mu_i) &=&  \sum_i  p^\mathrm{pt}_i (T, \tilde\mu_i)~, \\
 n (T, \mu_i) &=& \sum_i \frac{n^\mathrm{pt}_i (T, \tilde\mu_i)}
{1 + \sum_j n^\mathrm{pt}_j v^\mathrm{ex}_{ij}}~.
\end{eqnarray*}
The excluded volume effectively takes into account the strong repulsive interaction which has its 
origin in the Pauli principle for the fermionic constituents of the nuclear clusters.
This will be discussed in more detail in the next section.

\begin{figure}[!h]
\resizebox{0.45\textwidth}{!}{%
\includegraphics{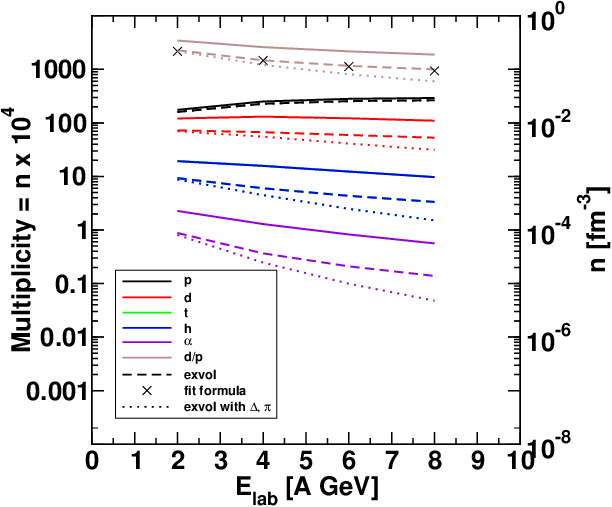}
}
\caption{Abundances of protons and of light clusters following LMA  for temperature and chemical potential values along the freeze-out line Eq.~(\ref{freezeout}) in Au+Au collisions in the NICA energy range 
(anticipating an energy scan with $E_{\rm lab}=2,~4,~6,~8$  A GeV).
In each case solid lines are for the pointlike particles, dashed for the excluded volume correction  for nucleons and clusters, and dotted ones including also the volume of pions and deltas. 
\label{lma-pauli}      
}
\end{figure}

Additionally, we consider pions ($m_\pi=137$ MeV, $r_\pi=0.8$ fm) and Delta baryons 
($m_\Delta=1232$ MeV, $r_\Delta=0.6$ fm) as particles which occupy volume.
The binding energies of the clusters (in MeV) are
$B_i= -2.22, -8.48, -7.72, -28.3$ for $i=d,t,h,\alpha$.
The results for the multiplicities of protons and light clusters according to the LMA without and with excluded volume correction are given in Fig.~\ref{lma-pauli} for thermodynamical parameters along the freeze-out curve in the $T-\mu_B$ diagram.
For comparison the results for the $d$ yield following from the fit formula \cite{Feckova:2016kjx} 
\begin{equation}
d/p = 0.8 \left[\sqrt{s_{NN}}/{\rm GeV} \right]^{-1.55} + 0.0036
\end{equation}
for the d/p ratio (crosses)  are also given.

\subsection{Quantum statistical light-cluster abundances}

A variety of issues tie in to the LMA description of yields, both pertaining to equilibrium and nonequilibrium aspects of the collisions.
Regarding equilibrium, medium effects may affect freeze-out yields 
giving results different from the simple LMA.
Regarding nonequilibrium, the LMA has nothing to say on why freeze-out occurs in the first place.
Further the LMA does not address spectra, occurrence of flow or possibility of $T$ and $\mu$ being different for different parts of the system at freeze-out.

Regarding dynamical complexities, a dynamical description with cluster production is needed, to be discussed in the next subsection.
Regarding equilibrium, a quantum statistical approach \cite{R15} based on the Greens function technique can be used to include mean-field effects and Pauli blocking to evaluate light cluster abundances. 
The extrapolation to high temperatures as shown in Fig.~\ref{PD-Mott} needs more investigations.
An alternative semi-empirical approach is the concept of an excluded volume as discussed above.

The effects of in-medium modifications on the EoS are investigated in heavy-ion collisions (HIC) near the Fermi energy \cite{Natowitz,NatowitzEoS}.
Clear signatures have been observed for the in-medium effects; 
an EoS based on the LMA (also denoted as nuclear statistical equilibrium = NSE) is not valid for the conditions considered there since it neglects the interaction of the clusters with the medium.

Investigations at NICA can thus test the validity of a simple LMA and to show the medium effects.
For this, these effects (self-energy and Pauli blocking) at high excitations have to be evaluated. 
Mott lines and changes in the cluster abundances and energy spectra are expected. 
Extrapolations of the low-temperature region to higher temperatures shown for the Mott densities in 
Fig.~\ref{PD-Mott} are probably subject to change. 
NICA being designed to cover the region of high densities is well disposed to investigate
the medium modification of light cluster particles.

An important issue is how medium effects modify the simple LMA.
The success of this simple LMA raises questions about the applicability of a freeze-out concept 
with a well-defined freeze out temperature $T_{\rm freeze}$
and density $\rho_{\rm freeze}$. 
A more appropriate description of the nonequilibrium process of HIC using transport codes 
will be indicated in the following section.

It is amazing that the simple LMA already describes the measured cluster multiplicities in good approximation. 
As seen from Fig.~\ref{PD-Mott}, conditions at NICA are close to the Mott lines where the bound states begin to be dissolved by medium effects. 
An appropriate approach to equilibrium is indispensable to analyze the cluster yields if the simple LMA ignoring all interactions is invalid. 
NICA is expected to be able to investigate matter at conditions where medium effects produced by interaction are relevant.

\subsection{Transport codes describing formation of light clusters}

Development of transport with cluster production has been a challenge.
Formal derivations invoked time-dependent Greens function formalism and solution of the von Neumann
equation for the statistical operator. 
Within a quasiparticle approach, coupled quantum kinetic equations can be derived which describe the 
time evolution of the Wigner  distribution functions for nucleons and light clusters \cite{Pawel,Kuhrts}.

In-medium cross sections (reaction rates) are introduced, and antisymmetrization is taken into account by Pauli blocking of the phase space where the states are already occupied by the medium. 
Presently only clusters with $A\le 3$ have been implemented;
the inclusion of the $\alpha$-particle distribution function is important for a realistic description and is a challenge in this approach.

An alternative approach has been proposed by A. Ono within the antisymmetrzed molecular dynamics (AMD) approach [18], in which nucleons are represented in terms of quantal wave packets, which are antisymmeztrized with each other. Cluster production can occur in 2-body collisions, the transition rate is determined by the overlap of the colliding nucleon pair with the quantal cluster wave function. A third partner takes up the energy-momentum balance. Heavier clusers are formed from collisions of lighter clusters. It is found that cluster-cluster collisions are important for a realistic description.  

\begin{figure}[h]
   \includegraphics[width=0.45\textwidth]{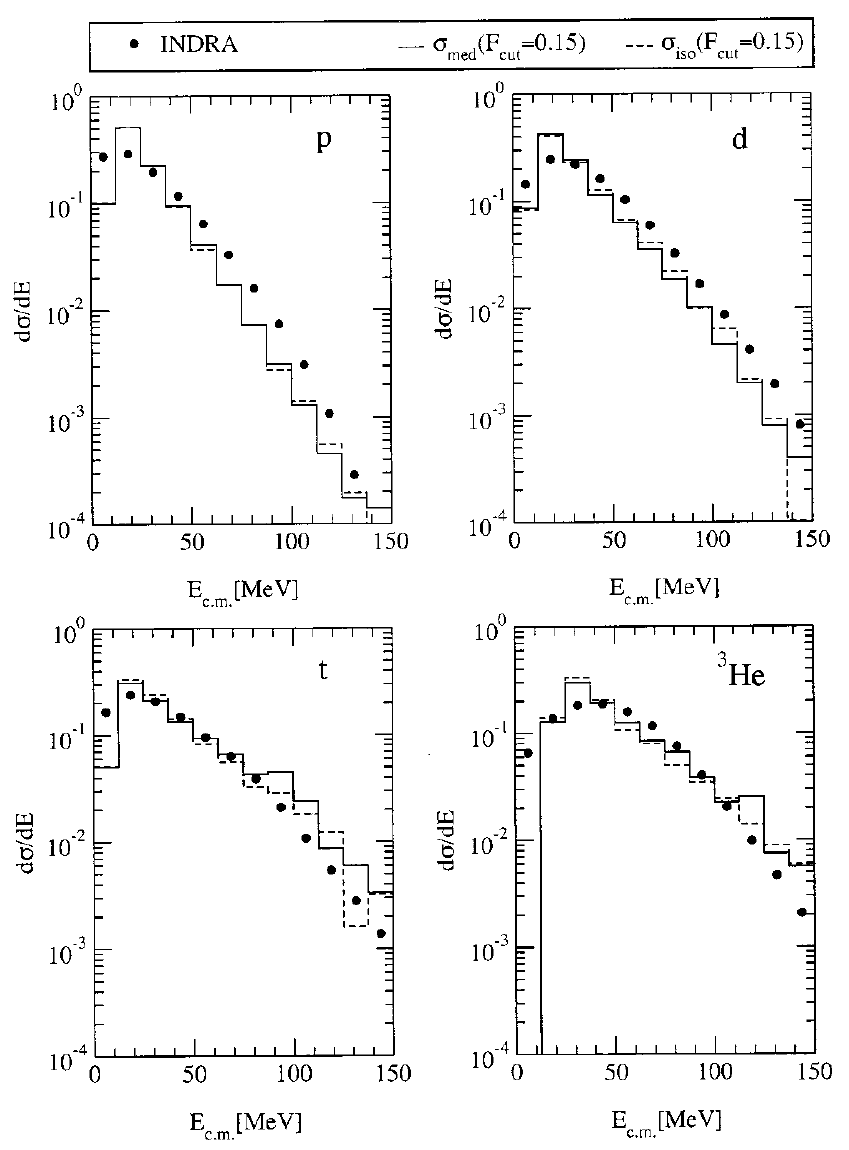}
    \caption{Differential cross sections for production of charged $A\le 3$ fragments in Xe+Se collisions at 50 MeV/nucleon.
    Results from the model of Ref.~\cite{Kuhrts} with cluster of mass $A\le 3$ as explicit degrees of freedom. 
    \label{Fig:Kuhrts}}    
 \end{figure}
\begin{figure}[h]
   \includegraphics[width=0.45\textwidth]{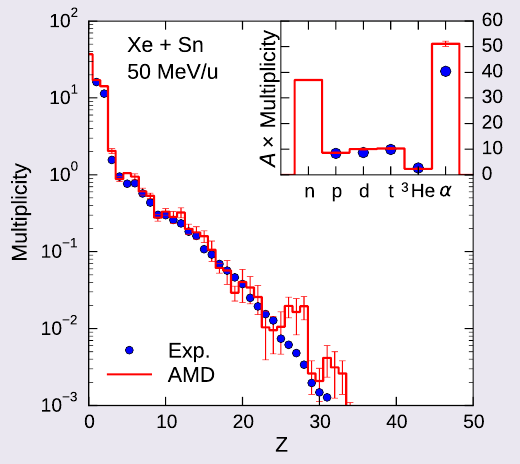}
    \caption{Multiplicity of different charged fragments in Xe+Sn collisions at 50 MeV/nucleon.
    Results from the AMD model of Ref.~\cite{Akira}, including also heavier cluster formation from 
    cluster-cluster collisions.
    \label{Fig:Akira}}    
 \end{figure}

Both of the discussed approaches have been applied to measurements performed at GANIL 
of central Xe + Sn reactions at 50 MeV/u, see Figs.~\ref{Fig:Kuhrts}, \ref{Fig:Akira}. 
Without any doubt, such an analysis is also necessary for the expected NICA data.

There is a variety of consistency issues when modeling the cluster production, e.g., between the two dynamic approaches discussed.
Do they predict the same yields?
How does the approach to the chemical equilibrium depend on included cluster formation processes?
When generating predictions for a system enclosed in a box, do the dynamic models yield results consistent with the LMA and/or more advanced equilibrium models?
Of general interest is the question of how well the dynamic models can incorporate in-medium effects such as the Mott effect.
In the work of Kuhrts et al.~\cite{Kuhrts} more sensitivity in the spectra was found to the effects of Pauli suppression of the clusters than to the reaction rates.

NICA experiments together with appropriate transport codes including cluster formation may give a better understanding of properties of dense matter.

\subsection{Three-fluid hydrodynamics simulations at NICA energies}

Three-fluid hydrodynamics (3FH) \cite{Ivanov:2005yw} was designed to simulate heavy-ion collisions at moderately relativistic energies. 
The 3-fluid approximation is a minimal way to simulate the finite
stopping power at the initial stage of the collision. 

At collision energies $E_{\rm{lab}} <$ 10$A$ GeV a substantial part of observed 
baryons is bound in light fragments (deutrons, tritons, $^3$He and $^4$He). 
Therefore there is a need to calculate the yield of these fragments, 
at least to properly predict the yield of protons. 
Light fragment formation in the 3FH model is
taken into account in terms of the coalescence model, which is similar
to that in  Appendix E of Ref.~\cite{Russkikh:1993ct}.  

\begin{figure}[!h]
\resizebox{0.4\textwidth}{!}{%
\includegraphics{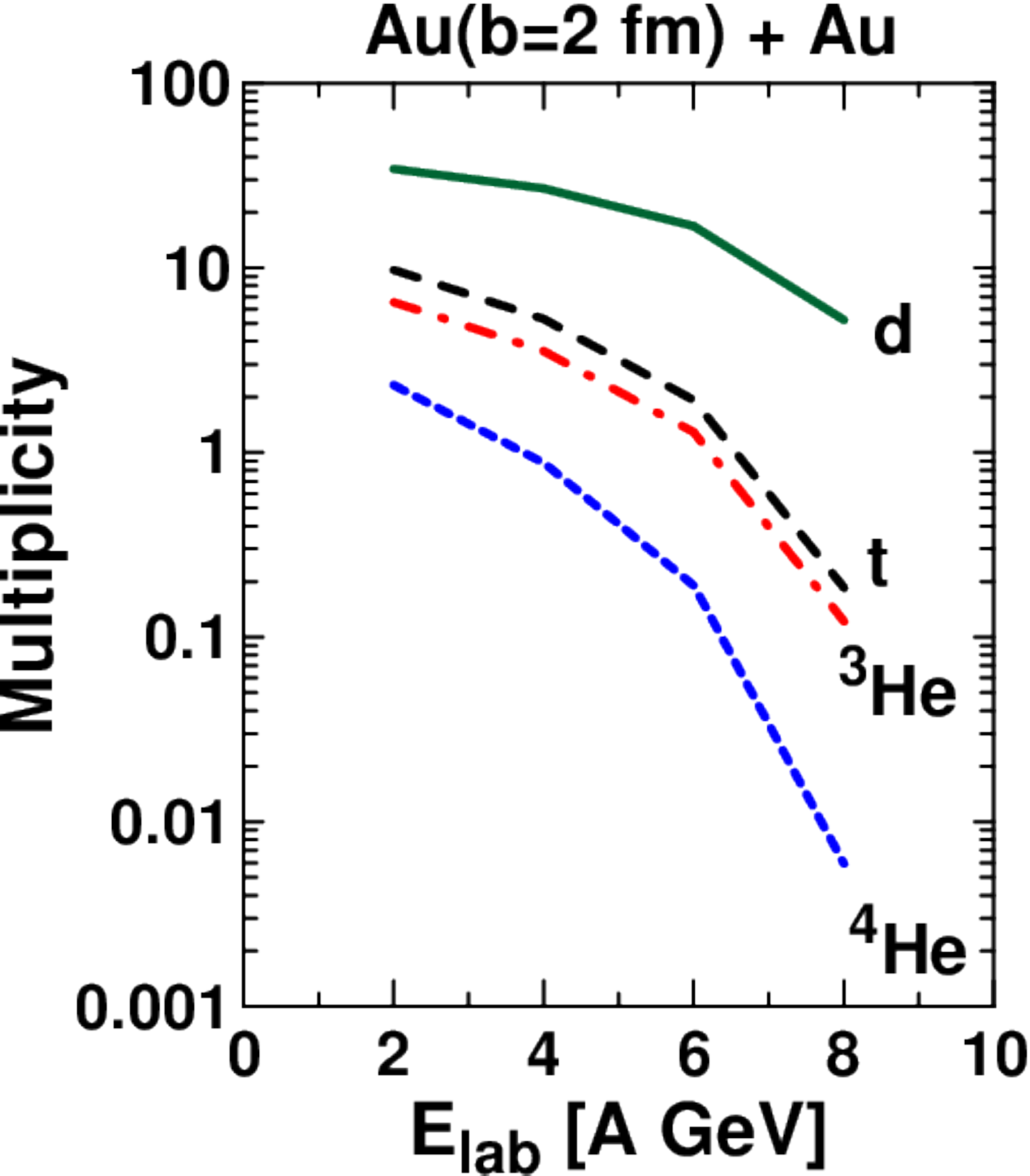}
}
\caption{Multiplicities of light clusters in central Au+Au collisions in the NICA energy range 
(calculated for an energy scan with $E_{\rm Lab}=2,~4,~6,~8$  A GeV). Results from a 3-fluid hydrodynamics description with cluster coalescence \cite{Russkikh:1993ct}.
\label{mult}      
}
\end{figure}

In Ref.~\cite{Russkikh:1993ct}, the coalescence coefficients were fitted at the
incident energy $0.8~A$ GeV. 
To extend this formulation to higher incident energies, only an overall scale of
the coalescence (the coefficient $C_{\rm{coal}}$ \cite{Ivanov:2005yw}) 
was applied, keeping ratios of yields of various fragments the same as in \cite{Russkikh:1993ct}. 
This was done because systematic experimental data on various fragments were lacking. 
The $C_{\rm{coal}}$ coefficients were chosen on the condition of the best 
reproduction of proton data. 
Figure \ref{mult} summarizes total multiplicities of various light fragments 
obtained within the 3FH model. This result is practically independent of the occurrence 
of a deconfinement transition in the EoS because the hadronic phase dominates at 
the considered energies. 

Recently, the relationship between the thermal and coalescence models for the production of light
nuclei in heavy-ion collisions has been discussed in Ref.~\cite{Mrowczynski:2016xqm}.

\section{Flow of light clusters}

\begin{figure}[!hb]
\resizebox{0.4\textwidth}{!}{%
\includegraphics{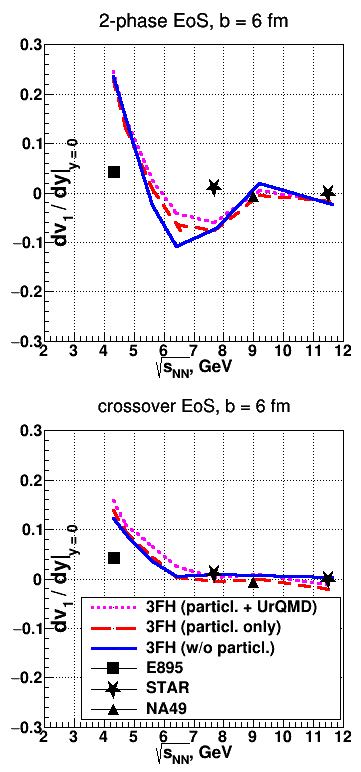}\\
}
\caption{Excitation function of the slope of the directed flow of protons in semicentral Au + Au 
collisions at impact parameter $b=6~$fm, from the three-fluid hydrodynamic code 
with freeze-out in the case of crossover (lower panel) and two-phase equations of state (upper panel), 
compared to data (symbols).
\label{fig:v1slope}
}
\end{figure}

Constraints for the cold, dense EoS as derived from flow measurements
\cite{Danielewicz:2002pu} have wide-ranging implications, from possibly 
signalling the deconfinement phase transition in compressed baryon matter produced
in heavy-ion collisions to consequences in nuclear 
astrophysics~\cite{Klahn:2006ir,Blaschke:2008cu,Klahn:2011au}.
Refinement of those constraints, both for zero and finite temperatures, filling
in gaps in systematics of flow anisotropies and combining the results from flow
with those from other observables, are urgent issues for upcoming experiments. 
Of particular urgency are the characteristics of the high-density phase 
transition and the location of the critical endpoint (CEP).

In this contribution, we suggest that the measurement of heavier
clusters (such as $d$, $t$, $h$, $\alpha$) may provide the necessary leverage 
for such a refinement.
Systematics of flow measurements by the FOPI Collaboration at GSI \cite{Rami:1998bf}
illustrate the principal strategy for the refinement, see also \cite{FOPI:2011aa}.
The energy ranges provided by the NICA facility, in fixed target and collider 
modes, are ideally suited for such an experimental programme, which we explain 
in the following in a bit more detail.

\begin{figure}[!hb]
\resizebox{0.45\textwidth}{!}{
\includegraphics{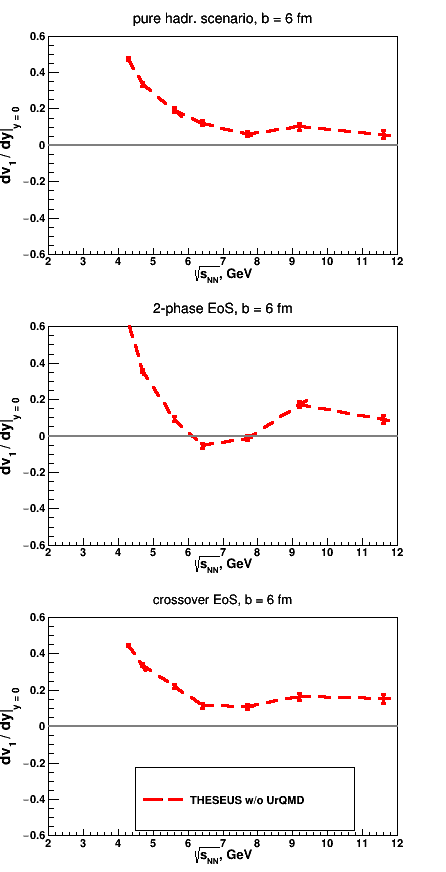}
}
\caption{Excitation function of the slope of the directed flow of deuterons in semicentral Au + Au 
collisions at impact parameter $b=6~$fm, from the three-fluid hydrodynamic code 
with freeze-out in three cases of EoS: purely hadronic (top panel), two-phase equation of state (middle panel) and  crossover (bottom panel).
\label{fig:d-flow}
}
\end{figure}

First and second order flow anisotropies are generated relatively early within
the collisions of heavy nuclei and, thus, can serve as useful probes of the
early pressure tensor in those collisions \cite{Sorge:1996pc,Shi:2001fm}.
The phase transition to the quark gluon plasma is generally expected to
produce a non-monotonous behavior of derivatives in the equilibrium pressure
(e.g., a dip in the speed of sound $c_s=\sqrt{d p/d \varepsilon}$)  depending on the  
early compression and temperature \cite{Teaney:2000cw}.
With different beam energies giving rise to different early compression and
temperature, the transition might be identified through a discontinuous
behavior of the excitation functions for the flow anisotropies 
\cite{Danielewicz:1998vz}.
The significance of such a signal could be augmented by finding a parallel
discontinuity in the excitation functions for the low-momentum correlation`
functions or for pion yields, both testing entropy \cite{Rischke:1996em}.

The above situation is, however, likely complicated by the fact that the
observables do not probe instantaneous local pressure but rather its effects
integrated over some volume and the evolution of a reaction.
To add to this, the transition to quark matter, which is expected to be crossed in the collisions
at NICA, may be weak and might not even be strictly a phase transition in the
thermodynamic sense.
In particular,  energy scan programs of heavy-ion collision
experiments such as at NICA probe the vicinity of the CEP
of first order phase transitions.
Therefore,  the baryon number or energy density as an order parameter change 
either continuously (at the crossover side of the CEP)
or with a very small jump (at the first order side of the CEP).
For this reason it is a difficult task to define easily accessible
observables which indicate the observation of the phase transition
and the relative position to the CEP.
Higher order derivatives of particle distributions (susceptibilities) have
been suggested as sensible quantities for identifying a phase transition
\cite{Friman:2011pf}.
They require, however, a very high statistics of identified hadrons in order
to derive these quantities with sufficient precision from the experiment.

Indeed, if the transition were strong, by now it would have likely been 
indicated by a significant discontinuity in the pion yield excitation function.
Given the opposite, the identification of quite subtle signs of discontinuities
in the excitation functions becomes a more important task.
In order to succeed in isolating the transition under the previously mentioned 
circumstances, it is  necessary to account as accurately as
possible for the effects of the early pressure at any one beam energy.

It has been suggested that the curvature of the proton rapidity
distribution at midrapidity might serve as a signal for a phase transition
\cite{Ivanov:2010cu}.
Protons being heavier than pions generally exhibit stronger emission
anisotropies.  
At any one velocity vector of a particle, on account of thermal motion, pions
stem from a wider range of freeze-out positions than do protons.
The averaging over emission positions suppresses the impact of collective
expansion velocities tied to specific locations in space.

Nuclear clusters, however, stem from an even more narrow region in space, than
do protons, and their velocity distribution generally better reflects the
distribution of collective velocities, with better emphasized anisotropies of
the collective flow.

In order to illustrate this fact, we show in Fig.~\ref{fig:v1slope} the excitation function 
of the slope of the directed flow of protons in semicentral Au+Au collisions.
We compare results of 3FH based simulations \cite{THESEUS} 
for an EoS with a first order phase transition (upper panel) with those for a crossover 
transition (lower panel). 
The first order phase transition reveals itself by a dip into negative slope values (antiflow)
for energies around $\sqrt{s_{NN}}\sim 6-7$ GeV while the crossover EoS shows in the same 
energy range a monotonic behaviour.

In Fig.~\ref{fig:d-flow}, we show the same observable as in  Fig.~\ref{fig:v1slope} for deuterons.
As in the case for protons, there is an antiflow for $\sqrt{s_{NN}}\sim 6-7$ GeV only for the case of the 
2-phase EoS while two alternative EoS without a first order phase transition do not exhibit the antiflow. 

Given the above, to increase prospects of identifying the transition through
collective flow anisotropies, it is utterly important to have detectors at
NICA capable of measuring clusters, especially $d$, $t$, $h$ and $\alpha$, and
this over a wide rapidity range.
In addition to helping with the flow, measurements of clusters would improve
precision in assessments of entropy \cite{Pal:2003rz} again tied to the 
transition.

\section{Conclusion}
From this contribution we suggest issues to be solved by the NICA project:
\begin{itemize}
\item Is the  ideal LMA applicable to describe the cluster yields, or are deviations observed?
This was shown at lower energies in the paper of Qin et al.  \cite{NatowitzEoS} which clearly indicates that  the ideal LMA cannot describe the data. Medium modifications are relevant.
\item Is the excluded volume an acceptable concept? Is quantum statistics a more appropriate systematic approach? Are signatures of the Mott points seen from the experiments?
\item Consequences of in-medium effects for the energy spectra of the light elements are seen in a suppression at low momenta
\cite{Ropke:1985dr}. 
Therefore, detectors should be designed to work in this region in order to investigate the effects of medium modifications on the energy spectra of light elements. 
\end{itemize}
In concluding our contribution, we summarize as follows.
Light clusters ($d, t, h, \alpha$) can be important indicators for the properties of matter produced
in HIC at NICA. Compared with the spectra of single nucleons, they probe the phase space more locally. The multiplicities and energy spectra, but also flow, contain signatures of matter in the early stages 
of the collision.  Transport theories including cluster production should be further developed to interpret the data.

\subsection*{Acknowledgement}
N.-U.B., D.B. and Iu.K. acknowledge support by the Polish NCN under grant number 
UMO-2011/02/A/ST2/00306.
D.B. acknowledges support by the MEPhI Academic Excellence Project 
under contract No. 02.a03.21.0005.
P.D. acknowledges support from the US National Science Foundation under grant number PHY-1403906 and FUSTIPEN under US Department of Energy grant number DE-FG02-10ER41700.
H.H.W. acknowledges support  from the DFG cluster of excellence "Origin and structure of the universe".
%


\end{document}